\documentclass[11pt,superscriptaddress,aps,prd,preprint,showpacs]{revtex4}
\usepackage[dvips]{graphicx}
\usepackage{amsfonts}
\usepackage{slashed}

\newcommand{\be}{\begin{equation}}
\newcommand{\en}{\end{equation}}
\newcommand{\ba}{\begin{eqnarray}}
\newcommand{\ea}{\end{eqnarray}}

\newcommand{\RM}[1]{\mathrm{#1}}

\begin{document}

\title{Dynamical Lorentz symmetry breaking in a $4D$ massless four-fermion model}

\author{J. F. Assun\c c\~ao}
\author{T. Mariz}
\affiliation{Instituto de F\'\i sica, Universidade Federal de Alagoas,\\ 57072-900, Macei\'o, Alagoas, Brazil}
\email{jfassuncao,tmariz@fis.ufal.br}

\author{J. R. Nascimento}
\author{A. Yu. Petrov}
\affiliation{Departamento de F\'{\i}sica, Universidade Federal da Para\'{\i}ba,\\
 Caixa Postal 5008, 58051-970, Jo\~ao Pessoa, Para\'{\i}ba, Brazil}
\email{jroberto,petrov@fisica.ufpb.br}


\begin{abstract}
In this paper, we study the spontaneous Lorentz symmetry breaking for a four-dimensional massless four-fermion model. Our methodology is based on use of the rationalized propagator. We show that a bumblebee potential arises as a result of one-loop calculations and displays nontrivial minima. Also we demonstrate that a phase transition restoring Lorentz invariance can occur at a finite temperature.
\end{abstract}


\maketitle

In field theory, the possible ways to implement the breaking of a symmetry are the explicit one, the anomalous one, and the spontaneous one. The framework designed to parametrize Lorentz symmetry breaking is the Standard Model Extension (SME) \cite{Colladay:1996iz,Colladay:1998fq,Kostelecky:2003fs}. It is an effective field theory incorporating different small additive terms explicitly violating the Lorentz symmetry, which allows to produce measurable corrections to Standard Model. The anomalous Lorentz symmetry breaking, whose origin can also be considered as explicit, is of special importance if the space-time possesses a nontrivial topology, such as, one of space-time dimensions is compact, or there is a linear defect in the space-time. It was shown in \cite{Klink} that in this case the Carroll-Field-Jackiw (CFJ) term \cite{Jackiw:1999yp} naturally emerges. However, it has been argued by Kostelecky \cite{Kostelecky:2003fs} that explicit Lorentz violation, within the SME in curved space-times, leads to incompatibility of the Bianchi identities with the covariant conservation laws for the energy-momentum and spin-density tensors. It is natural to expect that the similar situation occurs with the anomalous Lorentz breaking.
Hence, the spontaneous Lorentz symmetry breaking seems to be the only consistent mechanism to generate Lorentz violation in the SME on the curved background. The essence of this mechanism consists in coupling of vector (or, in general, tensor) fields through potentials of special form, so that
at minima of the potentials these fields acquire nontrivial "vacuum expectation values", thus introducing privileged space-time directions. 
The simplest field theory model allowing for spontaneous Lorentz symmetry breaking is the bumblebee model \cite{Kostelecky:1989jp,Kostelecky:1989jw,Kostelecky:2000mm,Bertolami:2005bh,Altschul:2005mu}:
\begin{eqnarray}\label{BB}
{\cal L}_B &=& -\frac{1}{4}F_{\mu\nu} F^{\mu\nu} + \bar\psi(i\slashed{\partial}-m-e\slashed{B}\gamma_5)\psi-\frac{\lambda}{4}\left(B_\mu B^\mu-\beta^2\right)^2,
\end{eqnarray}
where $F_{\mu\nu}=\partial_\mu B_\nu - \partial_\nu B_\mu$. By shifting the bumblebee field $B_{\mu}$ around its non-trivial vacuum expectation value (VEV) $\left< B_{\mu}\right> = \beta_{\mu}$, by the rule $B_{\mu}\to \beta_{\mu}+A_{\mu}$, where is assumed that $\left< A_{\mu}\right> =0$, the Lagrangian above becomes
\begin{eqnarray}
{\cal L}_B &=& -\frac{1}{4}F_{\mu\nu} F^{\mu\nu} + \bar\psi(i\slashed{\partial}-m-e\slashed{A}\gamma_5-\slashed{b}\gamma_5)\psi-\frac{\lambda}{4}\left(A_\mu A^\mu+\frac{2}{e} A\cdot b\right)^2,
\end{eqnarray}
with $b_\mu = e\beta_\mu$. Thus, we observe that the spontaneous Lorentz violation in (\ref{BB}) has generated the term $\bar\psi\slashed{b}\gamma_5\psi$, belonging to the SME Lagrangian, in which $\bar\psi\gamma^{\mu}\gamma_5\psi$ violates CPT symmetry, while $b_{\mu}$ violates Lorentz symmetry. From now on, we will concentrate on the massless case, i.e., we put $m=0$.

Different issues related to the bumblebee model have been studied in a number of papers (see, e.g., Refs.~\cite{Gomes:2007mq,Gomes:2008jw,Seifert:2009gi,Maluf:2014dpa,Nascimento:2014vva,Hernaski:2014jsa,Maluf:2015hda,Escobar:2017fdi}). In this work, we will follow the idea originally proposed in \cite{CW} that quantum corrections can give origin to the spontaneous symmetry breaking, and show that the bumblebee potential can be dynamically induced through radiative corrections from a self-interacting massless fermion theory, given by Lagrangian
\begin{eqnarray}\label{Lag}
{\cal L}_0 &=& \bar\psi i\slashed{\partial}\psi - \frac{G}{2}(\bar\psi\gamma_\mu\gamma_5\psi)^2.
\end{eqnarray}
 It is convenient to introduce an auxiliary field $B_\mu$, in order to eliminate the term $(\bar\psi\gamma_\mu\gamma_5\psi)^2$, so that the above expression can be rewritten as
\begin{eqnarray}
\label{rewrite}
{\cal L} &=& {\cal L}_0 + \frac{g^2}{2} \left(B_\mu-\frac{e}{g^2}\bar\psi\gamma_\mu\gamma_5\psi\right)^2 \nonumber \\
&=& \frac{g^2}{2}B_\mu B^\mu + \bar\psi(i\slashed{\partial}-e\slashed{B}\gamma_5)\psi,
\end{eqnarray}
with $G=e^2/g^2$. In \cite{Gomes:2007mq} this problem was studied for massive fermion fields on the base of the perturbative approach, where the fermion propagator is expanded in series in the constant vector $b_\mu$. Now, besides considering massless fermions, the nonperturbative description will be carried out, which, as we will see,  allows to obtain the bumblebee potential in a very simple way.

 In order to obtain the effective action, and consequently the bumblebee effective potential, we start with the generating functional
\begin{eqnarray}
Z(\bar \eta,\,\eta) &=& \int DB_\mu D\psi D\bar\psi e^{i\int d^4x({\cal L}+\bar\eta\psi+\bar\psi\eta)}\nonumber\\
&=& \int DB_{\mu} e^{i\int d^4x \frac{g^2}{2}B_{\mu}B^{\mu}}\int D\psi D\bar\psi e^{i\int d^4x(\bar\psi S^{-1}\psi+\bar\eta\psi+\bar\psi\eta)},
\end{eqnarray}
where $S^{-1}=i\slashed{\partial}-e \slashed{B}\gamma_{5}$ is the operator describing the quadratic action.  Now, by considering the shift in the fermion fields, $\psi\rightarrow \psi-S\eta$ and $\bar{\psi}\rightarrow \bar{\psi}-\bar{\eta}S$, so that $\bar\psi S^{-1}\psi+\bar\eta\psi+\bar\psi\eta \rightarrow \bar\psi S^{-1}\psi-\bar\eta S \eta$, we obtain
\begin{eqnarray}
Z(\bar \eta,\,\eta) &=& \int DB_{\mu} e^{i\int d^4x \frac{g^2}{2}B_{\mu}B^{\mu}}\int D\psi D\bar\psi e^{i\int d^4x(\bar\psi S^{-1}\psi-\bar\eta S \eta)}.
\end{eqnarray}
Finally, by performing the fermion integration, we get
\be\label{GF}
Z(\bar \eta,\,\eta) = \int DB_\mu \exp\left(iS_\RM{eff}[B] - i\int d^4x\, \bar\eta\, S\, \eta \right),
\en
where the effective action is given by
\begin{equation}\label{1}
S_\RM{eff}[B] = \frac{g^2}{2} \int d^4x\, B_\mu B^\mu -i \RM{Tr} \ln(\slashed{p}-e\slashed{B}\gamma_5).
\end{equation}
The $\RM{Tr}$ symbol stands for the trace over Dirac matrices as well as for the integration in momentum or coordinate spaces. The matrix trace can be readily calculated, so that for the effective potential, we have
\be\label{Vef}
V_\RM{eff} = -\frac{g^2}{2}B_\mu B^\mu + i\,\RM{tr} \int\frac{d^4p}{(2\pi)^4}\, \ln(\slashed{p}-e\slashed{B}\gamma_5).
\en

The nontrivial minimum of this potential can be obtained as usual, from the condition of vanishing the first derivative of the potential: 
\be\label{DVef}
\frac{dV_\RM{eff}}{dB_\mu}\Big|_{e B_\mu=b_\mu} =  - \frac{g^2}e b^\mu - i\,\Pi^\mu = 0,
\en
where the one-loop tadpole amplitude is 
\begin{equation}\label{Tad}
\Pi^\mu = \RM{tr} \int\frac{d^4p}{(2\pi)^4} \frac{i}{\,\slashed{p}-\slashed{b}\gamma_5}(-ie) \gamma^\mu\gamma_5.
\end{equation}

To evaluate the tensor $\Pi^{\mu}$, we will employ the exact propagator and use the dimensional regularization, with 't Hooft-Veltmann prescription \cite{tHooft:1972tcz}. For this, we first extend the 4-dimensional spacetime to a $D$-dimensional one, so that $d^4p/(2\pi)^4$ goes to $\mu^{4-D}d^D\bar{p}/(2\pi)^D$, where $\mu$ is an arbitrary scale parameter with the mass dimension 1. In the following, similarly to \cite{tHooft:1972tcz}, we introduce the anticommutation relation (in $D$-dimensional space): $\{\bar{\gamma}^{\mu},\bar{\gamma}^{\nu}\}=2\bar{g}^{\mu\nu}$, with the contraction $\bar{g}_{\mu\nu} \bar{g}^{\mu\nu}=D$. Then, we split the $D$-dimensional Dirac matrices $\bar{\gamma}^{\mu}$ and the $D$-dimensional metric tensor $\bar{g}^{\mu\nu}$ into 4-dimensional parts and $(D-4)$-dimensional parts, i.e., $\bar{\gamma}^{\mu}=\gamma^{\mu}+\hat{\gamma}^{\mu}$ and $\bar{g}^{\mu\nu}=g^{\mu\nu}+\hat{g}^{\mu\nu}$, so that now the Dirac matrices satisfy the relations
\begin{eqnarray}
\{\gamma^{\mu},\gamma^{\nu}\}=2g^{\mu\nu}\,, \{\hat{\gamma}^{\mu},\hat{\gamma}^{\nu}\}=2\hat{g}^{\mu\nu}\,, \{\gamma^{\mu},\hat{\gamma}^{\nu}\}=0,
\end{eqnarray}
and consequently the metric tensors have the contractions $g_{\mu\nu}g^{\mu\nu}=4$, $\hat{g}_{\mu\nu}\hat{g}^{\mu\nu}=D-4$, and $g_{\mu\nu}\hat{g}^{\mu\nu}=0$. We note that the most significant change found within this regularization is the introduction of the commutation relation
\begin{eqnarray}\label{}
[\hat{\gamma}^{\mu},\gamma^{5}]=0
\end{eqnarray}
and the maintenance of the anticommutation relation
\begin{eqnarray}
\{\gamma^{\mu},\gamma^{5}\}=0.
\end{eqnarray}

Following this approach, for the rationalization of the propagator $iG_{b}(p)=i(\bar{\slashed{p}}-\slashed{b}\gamma_5)^{-1}$, we use the one presented in Ref.~\cite{Assuncao:2015lfa}, given by
\begin{equation}\label{Gb}
G_b(p) = \frac{\bar p^2+b^2+2(\bar p \cdot b)\gamma_5+[\hat{\slashed{p}},\slashed{b}]\gamma_5}{(\bar p-b)^2(\bar p+b)^2-4\hat p^2b^2}(\bar{\slashed{p}}+\slashed{b}\gamma_5),
\end{equation}
with $\bar\slashed{p}=\bar p_\mu \bar\gamma^\mu$ and $\bar p_\mu=p_\mu+\hat p_\mu$, where we have taken into account that $\hat p_\mu\gamma^\mu=0=p_\mu\hat\gamma^\mu$ and $\hat p\cdot b=0=\hat p\cdot p$. However, it is more convenient to present the expansion of the above expression in terms of $\hat{p}^2$, i.e.,
\begin{equation}\label{Gb2}
G_b(p) = S_b(p)+\frac{4\hat p^2b^2}{(\bar p-b)^2(\bar p+b)^2}S_b(p) + \cdots,
\end{equation}
where
\begin{eqnarray}\label{Sb}
S_{b}(p) &=& \frac{\bar p^2+b^2+2(\bar p \cdot b)\gamma_5+[\hat{\slashed{p}},\slashed{b}]\gamma_5}{(\bar p-b)^2(\bar p+b)^2}(\bar{\slashed{p}}+\slashed{b}\gamma_5).
\end{eqnarray}
It is not difficult to see that the second and other higher order terms of the propagator~($\ref{Gb2}$), by power counting, yield finite contributions to the tadpole tensor~($\ref{Tad}$). Therefore, they can be disregarded after carrying out the contraction $\hat{g}_{\mu\nu}\hat{g}^{\mu\nu}=D-4$, with taking the $D\rightarrow 4$.

Thus, let us calculate Eq.~($\ref{Tad}$), with use of the propagator (\ref{Sb}). In order to perform the integrations, we first employ the Feynman parametrization. As a result, we have
\begin{equation}\label{Tad1}
\Pi^\mu = e\mu^{4-D}\int_{0}^{1} dx\ \RM{tr} \int \frac{d^D\bar{p}}{(2\pi)^D}\frac{(\bar{q}^2+b^2+2(\bar{q}\cdot b)\gamma_{5}+2\hat{\slashed{p}}\slashed{b}\gamma_{5})(\bar{\slashed{q}}+\slashed{b}\gamma_{5})\gamma^{\mu}\gamma_{5}}{\left(\bar{p}^2-M^2\right)^2},
\end{equation}
where $\bar{q}_{\mu}= \bar{p}_{\mu}+(2x-1)b_{\mu}$ and $M^2 = 4b^2(x-1)x$. Then, after the calculation of the trace, we obtain
\begin{equation}\label{Tad2}
\Pi^\mu = -4e\mu^{4-D}\int_{0}^{1} dx\ \int \frac{d^D\bar{p}}{(2\pi)^D}\frac{(\bar p^2-M^2-2\hat p^2)b^\mu-2\bar p^\mu (\bar p\cdot b)}{(\bar p^2-M^2)^2}.
\end{equation}
Now, after we integrate over the momentum $\bar p$ and Feynman parameter $x$, we get
\begin{eqnarray}
\Pi^{\mu} &=& \frac{i(D-4)\mu^{4-D}\pi^{1-\frac{D}{2}} (b^{2})^{\frac{D}{2}-1}\csc\left(\frac{\pi D}{2}\right) \Gamma\left(\frac{D}{2}\right)}{\Gamma(D)}b^{\mu}.
\end{eqnarray}
We note that the above result turns out to be finite in $D$ dimensions, which has a removable singularity in $D=4$, since $\lim\limits_{D\rightarrow 4} \left[(D-4)\csc \left(\frac{\pi  D}{2}\right)\right] = \frac{2}{\pi}$ (the arising of removable singularities is known to be characteristic for quantum corrections in Lorentz-breaking theories, see \cite{JackAmb} for the discussion). Therefore, we obtain
\begin{equation}\label{Pimu}
\Pi^\mu = \frac{ieb^2}{3\pi^2}b^\mu,
\end{equation}
so that the gap equation (\ref{DVef}) can be rewritten as 
\be\label{DVef2}
\frac{dV_\RM{eff}}{dB_\mu}\Big|_{eB_{\mu}=b_{\mu}} = \left(-\frac{1}{G}+\frac{b^2}{3\pi^2}\right)eb_\mu= 0,
\en
whose nontrivial solution ($b_{\mu} \neq 0$) is $b^2 = \frac{3\pi^2}{G}$,  with $G>0$ ($G<0$) for timelike (spacelike) $b_\mu$. The above expression (\ref{DVef2}) can be integrated, yielding the potential
\begin{equation}
\label{eq23}
V_\RM{eff} = -\frac{e^2b^2}{6\pi^2}B^2+\frac{e^4}{12\pi^2}B^4+\alpha,
\end{equation}
where $\alpha$ is a some constant. By choosing $\alpha=\frac{b^4}{12\pi^2}$, we have exactly the bumblebee potential of~(\ref{BB}), with $\lambda=\frac{e^4}{3\pi^2}$. Effectively, we have showed here that the bumblebee potential possessing nontrivial minima indeed can arise as a quantum correction in the four-fermion model. In other words, we have explicitly demonstrated that in our theory, the dynamical Lorentz symmetry breaking is possible at zero temperature. The next step consists in verifying of this possibility at the finite temperature.

From recent works \cite{Assuncao:2015lfa,Zyuzin:2012vn,Goswami:2012db}, it is known that in Lorentz-violating theory, thermal effects imply the suppression of temporal component of $b_\mu$, leading to the restoration of parity symmetry. In our case, we expect to arise a well-defined transition from a parity-breaking phase to a parity-symmetric one. In order to obtain the critical temperature corresponding to the restoring of parity symmetry, let us assume from now on that the system is in thermal equilibrium with a temperature $T=\beta^{-1}$. So, we transform the Eq.~(\ref{Tad2}) from Minkowski space to Euclidean one and split the internal momentum $\bar p^\mu$ in its  spatial and temporal components, performing the following replacements: $\bar{g}^{\mu\nu}\to-\bar{\delta}^{\mu\nu}$, i.e.,  $\bar{p}^2\to-\bar{p}^2$, $\hat{p}^2\to-\hat{p}^2$, $\bar p\cdot b\to-\bar p\cdot b$, $\bar p^\mu\to-\bar p^\mu$, and $b^\mu\to-b^\mu$, as well as 
\begin{equation}
\mu^{4-D}\int \frac{d^D \bar{p}}{(2\pi)^D} \to \mu^{3-d}\int \frac{d^d\vec p}{(2\pi)^d} \,i\int \frac{dp_0}{2\pi},
\end{equation}
and $\bar{p}^\mu=\vec{p}^\mu + p_0 u^\mu$, where $\vec p^\mu=(0,\vec p)$ and $u^\mu=(1,0,0,0)$, with $D=d+1$.

In addition, in thermal regime the antiperiodic boundary conditions for fermions lead to discrete values of $p_0$, i.e., $p_0 = (2n+1)\frac{\pi}{\beta}$, with $n$ being integer, so that ${\textstyle\int}\frac{dp_0}{2\pi}\rightarrow \frac{1}{\beta}{\textstyle\sum}_n$. Thus, we get
\begin{equation}\label{Tad3}
\Pi^\mu = -4ie\mu^{3-d}\int_{0}^{1} dx\ \frac1\beta\sum_n\int\frac{d^d\vec{p}}{(2\pi)^d}\frac{(\vec p^2+p_0^2+M^2-2\frac{\vec p^2}{d}(d-2))b^\mu+2(\frac{\vec p^2}{d}-p_0^2)(b\cdot u)u^\mu}{(\vec p^2+p_0^2+M^2)^2},
\end{equation}
where we have considered $\vec p_\alpha \vec p_\beta \to \frac{\vec p^2}{d}(\bar\delta_{\alpha\beta}-u_\alpha u_\beta)$ and taken into account $u_\mu \hat \delta^{\mu\nu}=0$. After we perform the momentum integration and carry out the sum (see \cite{Ford:1979ds}), we obtain
\begin{equation}
\Pi^\mu = -\int_0^1 dx \frac{2ie(x-1)x}{\pi^2}b^2b^{\mu}+4ieT^2\int_0^1 dx\int_{|\xi|}^\infty dz \frac{2z^2-\xi^2}{(z^2-\xi^2)^{1/2}}(1-\tanh(\pi z)) (b\cdot u)u^{\mu},
\end{equation}
with $\xi=\frac{M}{2\pi T}$. In the limit of high temperature (or also in the case of $b^2\ll T^2$), $\xi \to 0$, so that the above expression becomes
\begin{equation}
\Pi^\mu = \frac{i e b^2}{3\pi^2}b^{\mu}+\frac{ie T^2}{3}(b\cdot u)u^{\mu}.
\end{equation}

Thus, by considering also $b_\mu = \vec b_\mu + b_0 u^\mu$, where $\vec b^\mu=(0,\vec b)$, the gap equation (\ref{DVef}) can once more be rewritten as
\be\label{DVef2a}
\frac{dV_\RM{eff}}{dB_\mu}\Big|_{eB_{\mu}=b_{\mu}} = \left(-\frac{1}{G}+\frac{b^2}{3\pi^2}\right)e\vec b_\mu + \left(-\frac{1}{G}+\frac{T^2}{3}+\frac{b^2}{3\pi^2}\right)e b_0 u_\mu= 0.
\en
From this we see that the change of the behavior due to the temperature effects is seen only for its temporal component. The critical temperature above is defined in such a way that, when it is overcome, the effective potential does not exhibit nontrivial minima more, i.e., in $\left(-\frac{1}{G}+\frac{T^2}{3}+\frac{b^2}{3\pi^2}\right) = 0$ (for a purely timelike $b_0\ne0$), that is, $\frac{b^2}{3\pi^2}=\frac{1}{G}-\frac{T^2}{3}$, we have $\frac{1}{G}-\frac{T^2}{3}\le0$, so that $b_0$ must be zero. Therefore, the restoration of parity symmetry occurs at $T_c =\sqrt{\frac3G}$, with $G>0$ (one should remind that the consistent case corresponds to positive $G$, see the definition of $G$ used in (\ref{rewrite})), i.e., when the critical temperature is overcome, the $b_0$ should be imaginary and thus non-physical. We note also that if the $b_{\mu}$ is spacelike, with $b_0=0$, the temperature dependence completely disappears and there is no phase transitions.

Let us now study the dynamics of the bumblebee field $B_{\mu}$, around the nontrivial vacuum $\left< B_{\mu}\right> = \frac{b_\mu}{e}$, as we have seen,  
by considering $B_{\mu}\to \frac{b_\mu}{e}+A_{\mu}$, with $\left< A_{\mu}\right> =0$. For this, the generating functional (\ref{GF}) must be expressed in terms of the shifted field, i.e.,
\be\label{Z}
Z(\bar \eta,\,\eta) = \int DA_\mu \exp\left[iS_\RM{eff}[A] + i\int d^4x \left(\bar\eta\frac{1}{\slashed{p}-\slashed{b}\gamma_5-e\slashed{A}\gamma_5}\eta \right) \right],
\en
 where the effective action is  now given by
\begin{equation}
S_\RM{eff}[A] = \int d^4x\left(\frac{g^2}{2}A_\mu A^\mu+\frac{g^2}{e}A_\mu b^\mu+\frac{g^2}{2e^2}b_\mu b^\mu\right)-i \RM{Tr} \ln(\slashed{p}-\slashed{b}\gamma_5-e\slashed{A}\gamma_5).
\end{equation}
Up to field independent factors, which can be absorbed in the normalization of (\ref{Z}), we get
\begin{equation}
S'_\RM{eff}[A,b] = \int d^4x\left(\frac{g^2}{2}A_\mu A^\mu+\frac{g^2}{e}A_\mu b^\mu\right) + S^{(n)}_\RM{eff}[A],
\end{equation}
where
\begin{equation}\label{series}
S^{(n)}_\RM{eff}[A] = i \RM{Tr} \sum_{n=1}^\infty\frac1n\left[\frac i{\slashed{p}-\slashed{b}\gamma_5}(-ie)\slashed{A}\gamma_5\right]^n.
\end{equation}
The tensors which will be evaluated are the tadpole, the self-energy, the three and four point vertex functions of the  field $A_\mu $. These contributions present superficial divergences, but only the self-energy is really divergent. Initially, for $n=1$, we have
\begin{eqnarray}
S_\RM{eff}^{(1)}[A] &=& i \RM{Tr} \frac i{\slashed{p}-\slashed{b}\gamma_5}(-ie)\slashed{A}\gamma_5 =i\int d^4x\, \Pi^\mu A_\mu,
\end{eqnarray}
where $\Pi^\mu$ is the tadpole, given by Eq.~(\ref{Pimu}), cf.~(\ref{Tad}).

The self-energy contribution, which corresponds to $n=2$, takes the form
\begin{eqnarray}
S_\RM{eff}^{(2)}[A] &=& \frac i2 \RM{Tr} \frac i{\slashed{p}-\slashed{b}\gamma_5}(-ie)\slashed{A}\gamma_5 \frac i{\slashed{p}-\slashed{b}\gamma_5}(-ie)\slashed{A}\gamma_5 = \frac i2 \int d^4x\, \Pi^{\mu\nu} A_\mu A_\nu,
\end{eqnarray}
where
\begin{equation}\label{Pimunu0}
\Pi^{\mu\nu} = \RM{tr} \int \frac{d^4p}{(2\pi)^4} \frac{i}{\,\slashed{p}-\slashed{b}\gamma_5}(-ie)\gamma^\mu\gamma_5 \frac{i}{\,\slashed{p}\,-\,i\slashed{\partial}-\slashed{b}\gamma_5}(-ie)\gamma^\nu\gamma_5.
\end{equation}
We will follow the same route chosen to proceed with the first tensor (\ref{Tad}), however, we now expect the arising of divergent contributions (similarly to the standard QED), from terms proportional to $g^{\mu\nu}$ and $\partial^{\mu}\partial^{\nu}$, as well as finite contributions, which can be exact (like $b^{\mu}b^{\nu}$) or ambiguous (CFJ term) ones, since the Lorentz and CPT symmetries are also broken. In a general case of $D$ dimensions, these contributions are given by
\begin{eqnarray}
\Pi^{\mu\nu} &=& A\partial^\mu\partial^\nu+B\Box g^{\mu\nu}+C \epsilon^{\mu\nu\lambda\rho}b_\lambda \partial_\rho+E b^{\mu}b^\nu+F b^2g^{\mu\nu},
\end{eqnarray}
where
\begin{eqnarray}
A&=&- \frac{i(D-2)(D^2-20D+24)\csc\left(\frac{\pi  D}{2}\right)\Gamma \left(\frac{D}{2}\right)}{96\pi^{(D-2)/2} \Gamma\left(D\right)}\left(\frac{b^2}{\mu^2}\right)^{\frac{D}{2}-2}, \\
B&=&- \frac{i(5D^4-71D^3+338D^2-680D+480)\csc\left(\frac{\pi D}{2}\right)\Gamma\left(\frac{D}{2}\right)}{192\pi^{(D-2)/2}\Gamma \left(D\right)}\left(\frac{b^2}{\mu^2}\right)^{\frac{D}{2}-2},
\end{eqnarray}
are the divergent ones, and
\begin{eqnarray}
C&=&\frac{i(D^2-7D+8)(D-4)\csc\left(\frac{\pi D}{2}\right) \Gamma\left(\frac{D}{2}\right)}{8\pi^{(D-2)/2}\Gamma (D)} \left(\frac{b^2}{\mu^2}\right)^{\frac{D}{2}-2}, \\
E&=&-\frac{i(D-8) (D-4) (D-2) \csc\left(\frac{\pi  D}{2}\right) \Gamma\left(\frac{D}{2}\right)}{4\pi^{(D-2)/2}\Gamma \left(D\right)} \left(\frac{b^2}{\mu^2}\right)^{\frac{D}{2}-2}, \\
F&=&\frac{i(D^2-5D+8)(D-4)\csc\left(\frac{\pi D}{2}\right)\Gamma\left(\frac{D}{2}\right)}{4\pi^{(D-2)/2}\Gamma \left(D\right)}\left(\frac{b^2}{\mu^2}\right)^{\frac{D}{2}-2},
\end{eqnarray}
are the finite ones. Now, by expanding the above expressions around $D=4$, we obtain
\begin{eqnarray}\label{RTP}
\Pi^{\mu\nu} &=& -\frac{ie^2}{6\pi^2\epsilon} (g^{\mu\nu}\Box-\partial^\mu \partial^\nu) - \frac{ie^2}{6\pi^2} \epsilon^{\mu\nu\lambda\rho}b_\lambda \partial_\rho - \frac{ie^2}{24\pi^2}\partial^\mu \partial^\nu+\frac{ie^2b^2}{3\pi^2} g^{\mu\nu}+\frac{2ie^2}{3\pi^2} b^\mu b^\nu \\ 
 &&+ \frac{ie^2}{24\pi^2}\left[2\ln\left(\frac{b^2}{\mu'^2}\right)-1\right] (g^{\mu\nu}\Box-\partial^\mu \partial^\nu), \nonumber
\end{eqnarray}
valid if higher orders in derivatives are neglected, with $\epsilon = 4-D$ and $\mu'^2=4\pi\mu^2e^{-\gamma}$.

One can immediately see that the divergent part corresponds to that one arising in the standard QED, since it is independent of the Lorentz-breaking axial vector $b^{\mu}$. In the context of  Weyl semimetals (similar to massless QED), the CFJ contribution was calculated also with use of the 't Hooft-Veltmann prescription in \cite{Assuncao:2015lfa}, with the coefficient turns out to be three times greater than in  Eq.~(\ref{RTP}). This difference arises due to the presence of $\gamma_5$ in the vertex $(-ie)\gamma^\mu\gamma_5$.

Another consequence of this vertex is the invalidation of the Furry theorem, so that, the odd-order terms of the series in Eq.~(\ref{series}) not necessarily must vanish. So, for $n = 3$, the expression (\ref{series}) is written as
\begin{eqnarray}
S_\RM{eff}^{(3)}[A] &=& \frac i3 \RM{Tr} \frac i{\slashed{p}-\slashed{b}\gamma_5}(-ie)\slashed{A}\gamma_5 \frac i{\slashed{p}-\slashed{b}\gamma_5}(-ie)\slashed{A}\gamma_5 \frac i{\slashed{p}-\slashed{b}\gamma_5}(-ie)\slashed{A}\gamma_5 \nonumber\\
&=& \frac i3 \int d^4x\, \Pi^{\mu\nu\rho} A_\mu A_\nu A_\rho,
\end{eqnarray}
with
\begin{eqnarray}\label{Pimnr}
\Pi^{\mu\nu\lambda} &=& \RM{tr} \int \frac{d^4p}{(2\pi)^4} \frac{i}{\,\slashed{p}-\slashed{b}\gamma_5}(-ie) \gamma^\mu\gamma_5 \frac{i}{\,\slashed{p}\,-\,i\slashed{\partial}-\slashed{b}\gamma_5}(-ie) \gamma^\nu\gamma_5 \nonumber\\ && \times \frac{i}{\,\slashed{p}\,-\,i\slashed{\partial}\,-\,i\slashed{\partial}'-\slashed{b}\gamma_5}(-ie) \gamma^\lambda\gamma_5,
\end{eqnarray}
where the derivatives $\slashed{\partial}$ and $\slashed{\partial}'$ act on $A_\mu$ and $A_\nu$, respectively. Although there are divergent contributions in the above three point function, after the calculation of the trace they all vanish. Thus, we get
\begin{eqnarray}
\Pi^{\mu\nu\lambda} &=& \frac{i(D-4)\pi^{1-\frac{D}{2}} \left(b^2\right)^{\frac{D}{2}-3} \csc\left(\frac{\pi D}{2}\right) \Gamma\left(\frac{D}{2}+1\right)}{64\Gamma(D+1)}\{2(D-14)(D-6)(D-4)(D-2) b^{\mu}b^{\nu}b^{\lambda}\nonumber\\
&&-b^2\{D[D(D^2-17D+104)-324]+336\} (b^{\mu}g^{\nu\lambda}+b^{\nu}g^{\lambda\mu}+b^{\lambda}g^{\mu\nu})\},
\end{eqnarray}
valid for $\Box/m^2\ll1$. In four dimensions, the $b_{\mu}b_{\nu}b_{\lambda}$ contribution vanishes,  then, we obtain
\begin{equation}\label{Pi3}
\Pi^{\mu\nu\rho} = \frac{ie^3}{3\pi^2}(b^{\mu}g^{\nu\lambda}+b^{\nu}g^{\lambda\mu}+b^{\lambda}g^{\mu\nu}).
\end{equation}

Superficially, the fourth term of the series in (\ref{series}) is logarithmically divergent, but it results in a finite expression, since the leading term is similar to that one in QED, where, as it is known, it is finite. Thus, for $n=4$, the expression (\ref{series}) gives
\begin{eqnarray}
S_\RM{eff}^{(4)}[A] &=& \frac i4 \RM{Tr} \frac i{\slashed{p}-\slashed{b}\gamma_5}(-ie)\slashed{A}\gamma_5 \frac i{\slashed{p}-\slashed{b}\gamma_5}(-ie)\slashed{A}\gamma_5 \frac i{\slashed{p}-\slashed{b}\gamma_5}(-ie)\slashed{A}\gamma_5\frac i{\slashed{p}-\slashed{b}\gamma_5}(-ie)\slashed{A}\gamma_5 \nonumber\\
&=& \frac i4 \int d^4x\, \Pi^{\mu\nu\lambda\rho} A_\mu A_\nu A_\lambda A_\rho,
\end{eqnarray}
where
\begin{eqnarray}\label{Pimnr1}
\Pi^{\mu\nu\lambda\rho} &=& \RM{tr} \int \frac{d^4p}{(2\pi)^4} \frac{i}{\,\slashed{p}-\slashed{b}\gamma_5}(-ie) \gamma^\mu\gamma_5 \frac{i}{\,\slashed{p}\,-\,i\slashed{\partial}-\slashed{b}\gamma_5}(-ie) \gamma^\nu\gamma_5 \nonumber\\ &\times& \frac{i}{\,\slashed{p}\,-\,i\slashed{\partial}\,-\,i\slashed{\partial}'-\slashed{b}\gamma_5}(-ie) \gamma^\lambda\gamma_5\frac{i}{\,\slashed{p}\,-\,i\slashed{\partial}\,-\,i\slashed{\partial}'-\,i\slashed{\partial}''-\slashed{b}\gamma_5}(-ie) \gamma^\rho\gamma_5.
\end{eqnarray}
In order to induce the bumblebee model, it is sufficient to single out the $g^{\mu\nu}g^{\lambda\rho}$ contribution, together with its permutations. The result is
\begin{eqnarray}
\Pi^{\mu\nu\lambda\rho} &=& \frac{i (D-4) \pi ^{1-\frac{D}{2}} \left(b^2\right)^{\frac{D}{2}-2} \csc \left(\frac{\pi  D}{2}\right) \Gamma \left(\frac{D}{2}+6\right)}{288 \left(D^3+3 D^2-D-3\right) \Gamma (D-2)}(g^{\mu \nu} g^{\lambda\rho}-g^{\mu\lambda} g^{\nu\rho }+g^{\mu\rho} g^{\nu\lambda}),
\end{eqnarray}
which, by expanding around $D=4$, yields
\begin{eqnarray}\label{Pi4}
\Pi^{\mu\nu\lambda\rho} &=& \frac{ie^4}{3\pi^2}(g^{\mu \nu} g^{\lambda\rho}-g^{\mu\lambda} g^{\nu\rho }+g^{\mu\rho} g^{\nu\lambda}).
\end{eqnarray}

Finally, by taking into account the results (\ref{Pimu}), (\ref{RTP}), (\ref{Pi3}), and (\ref{Pi4}), let us now write the one-loop contribution to the effective Lagrangian, as follows: 
\begin{eqnarray}\label{lag}
{\cal L} &=& -\frac{1}{4Z_3}F_{\mu\nu}F^{\mu\nu} + \frac{e^2}{24\pi^2}b^\mu\epsilon_{\mu\nu\lambda\rho}A^\nu F^{\lambda\rho} - \frac{e^2}{48\pi^2}(\partial_\mu A^\mu)^2 - \frac{e^4}{12\pi^2}\left(A_\mu A^\mu + \frac2e A\cdot b\right)^2 \nonumber\\
&&+\frac{e}{2b^2}A_\mu A^\mu \left\langle A_\nu \right\rangle b^\nu + \left\langle A_\mu \right\rangle A^\mu,
\end{eqnarray}
where
\begin{equation}\label{Z3}
\frac{1}{Z_3}= \frac{e^2}{6\pi^2\epsilon}-\frac{e^2}{12\pi^2}\left[\ln\left(\frac{b^2}{\mu'^2}\right)-\frac12\right]
\end{equation}
and
\begin{eqnarray}
\left\langle A_\mu \right\rangle = \left(\frac{1}{G}-\frac{b^2}{3\pi^2}\right)eb_\mu.
\end{eqnarray}
As obviously  $\left\langle A_\mu \right\rangle = 0$, because of the unique nontrivial solution (\ref{DVef2}), the above Lagrangian assumes the form
\be\label{L}
{\cal L} = -\frac14 F_{\RM{R}\mu\nu}F_\RM{R}^{\mu\nu} + \frac{e_\RM{R}^2}{24\pi^2}b^\mu\epsilon_{\mu\nu\lambda\rho}A_\RM{R}^\nu F_\RM{R}^{\lambda\rho} - \frac{e_\RM{R}^2}{48\pi^2}(\partial_\mu A_\RM{R}^\mu)^2 - \frac{e_\RM{R}^4}{12\pi^2}\left(A_{\RM{R}\mu} A_\RM{R}^\mu + \frac2{e_\RM{R}} A_\RM{R}\cdot b\right)^2,
\en
where we have considered the renormalized field $A_R^\mu=Z_3^{-1/2}A^\mu$, as well as the renormalized coupling constant $e_R=Z_3^{1/2}e$. This expression is exactly the Lagrangian of extended QED with the CFJ term, plus the gauge-fixing term, with the positively defined potential.

We have demonstrated explicitly that the bumblebee action can arise as a one-loop quantum correction. The key conclusion of our result is that our potential, looking like (see Eq.~(\ref{L}))
$$
V_\RM{eff}=\frac{e_\RM{R}^4}{12\pi^2}\left(A_{\RM{R}\mu} A_\RM{R}^\mu + \frac2{e_\RM{R}} A_\RM{R}\cdot b\right)^2,
$$
which is the potential (\ref{eq23}), with the shifted $B_{\mu}\to \frac{b_\mu}{e}+A_{\mu}$, is positively definite, which immediately implies that it possesses the minima. Therefore, we conclude with the statement that the effective potential in our theory is bounded from below, and, moreover, the theory possesses a set of minima, if the temperature is lower than the critical one, $T_c=\sqrt{\frac{3}{G}}$. When this temperature is overcame, the theory has only one minimum, and the parity symmetry is restored.  In~\cite{Gomes:2008jw,Gomes:2007mq} a similar study has been carried out, with use of different methodologies (massive fermions and expanded propagator), however, the positive definiteness of the potential was not discussed explicitly.

{\bf Acknowledgements.} This work was partially supported by Conselho
Nacional de Desenvolvimento Cient\'{\i}fico e Tecnol\'{o}gico (CNPq). The work by A. Yu. P. has been supported by the
CNPq project No. 303783/2015-0.

\end{document}